\documentclass[conference]{IEEEtran}

\IEEEoverridecommandlockouts    

\usepackage{amsmath,amssymb,amsfonts}
\usepackage{algorithmic}
\usepackage{graphicx}
\usepackage{textcomp}
\usepackage{xcolor}
\usepackage[mathscr]{euscript}

\usepackage{cite}
\usepackage{flushend}
\usepackage{tikz}
\usepackage{multirow}
\def\BibTeX{{\rm B\kern-.05em{\sc i\kern-.025em b}\kern-.08em
    T\kern-.1667em\lower.7ex\hbox{E}\kern-.125emX}}
\newcommand\T{\rule{0pt}{2.6ex}}

\title{A Deep Reinforcement Learning-based Reserve Optimization in Active Distribution Systems for Tertiary Frequency Regulation}

\author{Mukesh Gautam\textsuperscript{\textdagger}, \emph{Student Member, IEEE}, Rakib Hossain, \emph{Student Member, IEEE},\\ Mohammad MansourLakouraj, \emph{Student Member, IEEE}, Narayan Bhusal, \emph{Student Member, IEEE}, \\Mohammed Benidris, \emph{Senior Member, IEEE}, and Hanif Livani, \emph{Senior Member, IEEE}, \\ Department of Electrical \& Biomedical Engineering, University of Nevada, Reno, Reno, NV 89557 \\
(email: {}\textsuperscript{\textdagger}mukesh.gautam@nevada.unr.edu)\vspace{-0.2ex}}

\begin{document}
\maketitle
\begin{abstract}
Federal Energy Regulatory Commission (FERC) Orders 841 and 2222 have recommended that distributed energy resources (DERs) should participate in energy and reserve markets; therefore, a mechanism needs to be developed to facilitate DERs’ participation at the distribution level.
Although the available reserve from a single distribution system may not be sufficient for tertiary frequency regulation, stacked and coordinated contributions from several distribution systems can enable them participate in tertiary frequency regulation at scale. This paper proposes a deep reinforcement learning (DRL)-based approach for optimization of requested aggregated reserves by system operators among the clusters of DERs. The co-optimization of cost of reserve, distribution network loss, and voltage regulation of the feeders are considered while optimizing the reserves among participating DERs. The proposed framework adopts deep deterministic policy gradient (DDPG), which is an algorithm based on an actor-critic method. The effectiveness of the proposed method for allocating reserves among DERs is demonstrated through case studies on a modified IEEE 34-node distribution system. 
\end{abstract}

\section{Introduction}
Recently, large-scale integration of distributed energy resources (DERs) in power grids has seen a sustained growth rate and momentum \cite{mather2018going}. Although integration of DERs provides many opportunities and benefits for the grid, it has also brought new challenges for optimal and reliable operation of the grid including efficient frequency control and regulation. Frequency regulation problem (e.g., frequency deviation) occurs when there is an imbalance between the generation and load, which can happen due to several factors including faults, large load changes, generating units tripping, and islanding parts of the grid \cite{zhou2019optimal}. In such scenarios, frequency regulating schemes come into play to compensate for frequency deviations. Frequency regulation schemes are usually categorized into primary, secondary, and tertiary frequency regulation schemes. For tertiary frequency regulation, under vertically integrated monopolistic structure of utilities, system operators set operating points of individual generators based on security-constrained optimal power flow (SCOPF) solution to minimize the overall operating cost of the system subject to network and reserve constraints. On the other hand, in deregulated power systems, the main function of tertiary frequency regulation schemes is to maximize the net social welfare through allocating adequate reserve from generators or DERs participating in primary and secondary frequency regulation \cite{machowski2020power}. Although the contribution of a single DER in frequency control and regulation is not significant, the accumulated contribution from a fleet of DERs can enable them to collectively participate in frequency control and regulation. However, allocating impactful reserves from DERs is a challenging task and requires flexible and efficient solutions. 

Various analytical and data-driven approaches have been proposed in the literature for tertiary frequency regulation and reserve optimization in transmission systems. An approach for optimal tertiary frequency control has been proposed in \cite{perninge2017optimal}, which also considers regulation based on the electricity market.
In \cite{bovera2021data}, a data-driven approach for the estimation of secondary and tertiary reserve has been presented and tested in a real-life case study. A robust optimization approach has been proposed in \cite{zugno2015robust} for co-optimizing energy and reserve in electricity markets. 
Although several methods and algorithms have been developed and employed for tertiary frequency regulation and reserve optimization at transmission system level, allocating reserves from active distribution systems for the tertiary frequency regulation is still a challenge.

Learning-driven approaches have been used to tackle the limitations of analytical and population-based approaches since learning-driven approaches can handle uncertainties by extracting knowledge from historical data. Moreover, learning-driven models don't need to be solved whenever new scenarios are encountered because of their ability to use their knowledge gained from the historical data to solve for the new scenarios. Out of various learning-driven approaches, reinforcement learning (RL)-based approaches have the capabilities to learn from experiences during online operations \cite{zai2020deep, lillicrap2015continuous}. This paper proposes a deep reinforcement learning (DRL)-based framework for reserve optimization in active distribution systems for tertiary frequency regulation. The proposed framework aims to minimize reserve cost, power loss, and average voltage deviation. The deep deterministic policy gradient (DDPG), an algorithm based on the actor-critic method, has been adopted to solve the optimization problem. A mechanism called experience replay is used to allow batch training of the DDPG algorithm resulting in a stable training. In order to stabilize the update rule of the main actor and critic networks, target actor and critic networks are implemented. The actor network selects actions (here, scheduled reserves of each DER) based on the state (here, available reserves and reserve bid prices of each DER, and the total reserve requested by the system operator). In order to solve the exploitation-exploration dilemma, Gaussian noises are added while executing the actions. The DRL agent gets rewarded (or penalized) while executing the actions. The critic network evaluates the state and the action pairs using the mean squared error (MSE), which is a commonly employed loss function for regression. Finally, the target actor and critic network parameters are updated based on the soft-update rule. The proposed framework is validated through case studies on a modified IEEE 34-node distribution test system, and the results show that the proposed framework can effectively find optimal or near-optimal solutions. 

The rest of the paper is structured as follows. Section~\ref{model} describes the mathematical modeling of the reserve optimization problem under consideration. Section~\ref{proposed} explains about the proposed DRL framework. Section~\ref{cases} presents case studies and discussions. Section~\ref{conclusion} provides concluding remarks.

\section{Mathematical Modeling}\label{model}
This section presents the mathematical formulation of the reserve optimization problem and describes states, actions, and reward function in the context of the reserve optimization. 
\subsection{Problem Formulation}
This subsection presents the objective functions and the constraints of the reserve optimization problem.
\subsubsection{Objective Functions}
The reserve optimization problem is formulated for minimization of three objective functions as follows.

\textit{(a) Total Reserve Cost:} The reserve cost of an individual DER is the product of scheduled reserve amount and its reserve bid price. The total reserve cost (TRC) is the sum of reserve costs of individual DERs.
\begin{equation}
    TRC = \sum_{i=1}^{n} \pi_{i}R_i 
\end{equation}
where $n$ is the total number of DERs; $\pi_i$ is reserve bid price of the $i$th DER; and $R_i$ is scheduled reserve of the $i$th DER.

\textit{(b) Power Loss:} The total power loss of the system is calculated by adding power losses of all branches of the distribution network.
\begin{equation}
P_{loss} = \sum_{k \in \Phi_{B}} P_{loss,k} 
\end{equation}
where $\Phi_{B}$ is the set of branches (or lines); and $P_{loss,k}$ is the power loss of the $k$th branch.

\textit{(c) Average Voltage Deviation:} The voltage deviation at a node is the absolute value of the difference between the node voltage and the reference voltage. The average voltage deviation (AVD) is the average of the voltage deviations at all nodes of the system, which can be calculated as follows.
\begin{equation}
    AVD = \sum_{k=1}^{N} \frac{\vert V_k - V_{ref}\vert}{N} 
\end{equation}
where $N$ is the total number of nodes in the system; $V_k$ is the voltage of node $k$; and $V_{ref}$ is reference voltage (usually set equal to 1.0).
\subsubsection{Constraints}
The reserve optimization problem under consideration is subjected to various constraints including nodal power balance constraints, reserve allocation constraint, and reserve limit constraints.

\textit{(a) Node power balance constraints:} The power balance constraints at each node of the system can be expressed as follows. 
\begin{equation}
    \sum_{j\in \Phi_g(j)} P_{g,j} + \sum_{l\in \Phi_L(j)} P_{l,j} = P_{D,j} \label{eqn:Pbal}
\end{equation}
where $\Phi_g(j)$ is the set of sources (including DERs) connected to node $j$; $\Phi_L(j)$ is the set of lines connected to node $j$; $P_{g,j}$ is the power injected from source $j$; $P_{D,j}$ is the load at node $j$; and $P_{l,j}$ is the line power flow from node $l$ to node $j$.

\textit{(b) Reserve allocation constraint:} The sum total of the allocated reserves should be equal to the total reserve requested by the system operator.
\begin{equation}
    \sum_{i=1}^n R_i = R_{tot}
\end{equation}
where $R_{tot}$ is the total reserve requested by the system operator.

\textit{(c) Reserve limit constraint:} The scheduled reserve of each DER is limited by its maximum reserve availability as expressed in \eqref{eqn:Rlim}.
\begin{equation}
    R_i \leq R_i^{max}, \forall i \label{eqn:Rlim}
\end{equation}
where $R_i^{max}$ is the maximum reserve available from the $i$th DER.

\subsection{States, Actions, and Reward Function}
 For the reserve optimization problem under consideration, a vector of available reserve capacities of DERs, reserve bid prices of DERs, and the total reserve requested by the system operator is taken as the state. The action is a vector of scheduled reserves of DERs. The reward function is designed in a way to represent total reserve cost, reserve limit constraint, system power loss, and voltage regulation. The total reward at time step $t$ is computed as follows.
\begin{equation}
    R_t = -R_t^c - \beta_r R_t^r - \beta_p R_t^p + \beta_v R_t^v \label{eqn:Rwd_fcn}
\end{equation}
where $\beta_r$, $\beta_p$, and $\beta_v$ represent weighting factors which incentivize the RL agent to maximize the reward. 

The first term of the reward equation \eqref{eqn:Rwd_fcn}, i.e., $R_t^c$, represents the component of reward function related to reserve cost, and it encourages the RL agent to keep it as low as possible (a negative sign is given to this term). $R_t^c$ is calculated as follows.
\begin{equation}
    R_t^c = \frac{\sum_{i=1}^n \pi_i R_i}{R_{tot}}
\end{equation}

In the second term of the reward equation \eqref{eqn:Rwd_fcn}, $R_t^r$ represents the component of reward function related to reserve limit constraints, which is calculated as follows.
\begin{equation}
    R_t^r = n_r \sum_{m \in \Phi_r} R_i - R_i^{max}
\end{equation}
where $\Phi_r$ is the set of violated reserve limit constraints; and $n_r$ is cardinality of the set $\Phi_r$.

In the third term of the reward equation \eqref{eqn:Rwd_fcn}, $R_t^p$ represents the system power loss $P_{loss}$. In the fourth term of the reward equation \eqref{eqn:Rwd_fcn}, $R_t^v$ represents the component of reward function related to voltage regulation, and it encourages the RL agent to keep the bus voltages within limits and close to a predefined reference value, $V_{ref}$. $R_t^v$ is calculated as follows.
\begin{equation}
    R_t^v = \frac{1}{N} \sum_{k=1}^N r_t^k
\end{equation}
where $r_t^k$ represents the reward for voltage at the $k$th node. 
If voltage $V_t^k$ at the $k$th node at time step $t$ is between $V_{ref}$ and maximum voltage limit ($V_{ub}$), the reward is calculated as follows.
\begin{equation}
    r_t^{k} = \frac{V_{ub}-V_t^{k}}{V_{ub}-V_{ref}}
\end{equation}
If voltage $V_t^k$ at the $k$th node at time step $t$ is between minimum voltage limit ($V_{lb}$) and $V_{ref}$, the reward is calculated as follows.
\begin{equation}
    r_t^{k} = \frac{V_t^{k}-V_{lb}}{V_{ref}-V_{lb}}
\end{equation}
If the node voltage $V_t^k$ at time step $t$ is outside the limits, a negative reward (or penalty) is assigned to $r_t^k$.

\section{The Proposed DRL Framework}\label{proposed}
This work leverages recently advanced reinforcement learning techniques to solve reserve optimization problem in active distribution systems. This section provides a brief overview of reinforcement learning and training attributes of the proposed framework. 

\subsection{Reinforcement Learning and Deep Deterministic Policy Gradient (DDPG)}
A reinforcement learning (RL) is a branch of machine learning which consists of four main integrands: policy, reward, value functions, and environment model. An RL agent decides the action to be taken based on the policy. The policy maps states to actions. When the agent takes an action, it gets rewarded (or penalized). Value function calculates the expected value of cumulative reward that an agent gets when it follows a certain policy. There are different algorithms for RL. The choice of an algorithm depends on many factors such as the continuous/discrete nature of states, continuous/discrete action-space, etc. In this paper, both state-space and action-space are continuous in nature. Moreover, in order to allow the agents to make decisions based on unstructured input data, deep learning is incorporated into the solution. Therefore, DDPG is the best candidate algorithm for the problem under consideration, which is a model-free off-policy algorithm. The DDPG algorithm combines both Deterministic Policy Gradient (DPG) and Deep Q-Network (DQN). DDPG falls in the category of Actor-Critic method with two distinct networks: an actor and a critic. The actor network decides what to do based on current state and outputs action values. Contrary to stochastic policy gradient algorithms that output action probabilities, the actor network in case of DPG outputs action values, which are deterministic in nature. In order to allow the agent to explore possible actions, the noise is added to the output action values, which helps to unravel the exploitation-exploration dilemma. The critic network evaluates state and action pairs based on the reward the agent gets from the environment. 

During online training, when small batches of data are trained at a time, it leads to the corruption of the old information already learned by the agent \cite{zai2020deep}. This process of corruption of the old information is referred to as catastrophic forgetting. In order to mitigate catastrophic forgetting, a mechanism called experience replay is implemented, which allows batch training of RL algorithms resulting in a stable training. In addition, target networks, which are copies of the main networks, are implemented for both actor and critic to stabilize the update rule of the main networks \cite{zai2020deep}.

\subsection{Training Attributes}
The training of the proposed DRL framework is performed for a certain number of episodes. Since DDPG is based on actor-critic method, Q-function parameter ($\phi$) of the critic network and policy parameter ($\theta$) of the actor network are initialized. Target network parameters ($\phi_{targ}$ and $\theta_{targ}$) are also set equal to main network parameters. At the start of each episode, the state is initialized by resetting the environment. The resetting of the environment, here, implies getting a new state at the start of each episode. Each episode consists of a certain number of time steps. In each time step, the actor network chooses actions (scheduled reserves of DERs). The actions are implemented after adding the Gaussian noises. The following equation, referred to as the Bellman equation or Q-function, is utilized to compute the value of a particular deterministic policy ($\mu$).
\begin{equation}
    Q(S_t,A_t) = \mathbb{E}[R(S_t,A_t)+\gamma\times Q(S_{t+1},\mu(S_{t+1}))] \mbox{,} \label{eqn:Qval}
\end{equation}
where $\mathbb{E}$ denotes expectation operator; $R(S_t,A_t)$ denotes the reward function of the state $S_t$ and the action $A_t$; and $\gamma$ denotes the discounting factor.

Instead of iteratively updating the Q-function, the critic network is trained and its parameters are optimized to minimize the mean-squared error (MSE) loss function (i.e., regression loss function), which is expressed as follows \cite{lillicrap2015continuous}.
\begin{equation}
L(\phi)=\mathbb{E}[(Q(S_{t}, A_{t}\vert \phi)-y_t)^{2}]
\label{eqn:loss_fun}
\end{equation}
where $y_t$ denotes the target Q-function of the critic network, which is defined as follows.
\begin{equation}
    y_t = R(S_{t}, A_{t})-\gamma \times Q(S_{t+1},\mu(S_{t+1})\vert \phi) \label{eqn:targetQ}
\end{equation}

The policy of the actor network is updated by performing gradient ascent to solve the following function.
\begin{equation}
    J = \max\limits_{\theta} \mathbb{E}[Q(S_t, \mu(S_t|\theta)]
\end{equation}
where $\mu(S_t|\theta)$ is a parameterized actor function which maps states to actions in a deterministic manner.

The target actor and critic network parameters are updated with a small constant $\rho$ as follows.
\begin{equation}
\begin{aligned}
     \phi_{targ} \leftarrow \rho \phi_{targ} + (1-\rho) \phi \\
    \theta_{targ} \leftarrow \rho \theta_{targ} + (1-\rho) \theta   
\end{aligned}
\end{equation}

\section{Case Studies and Discussions}\label{cases}
\begin{figure}
    \centering
    \includegraphics[scale=0.55]{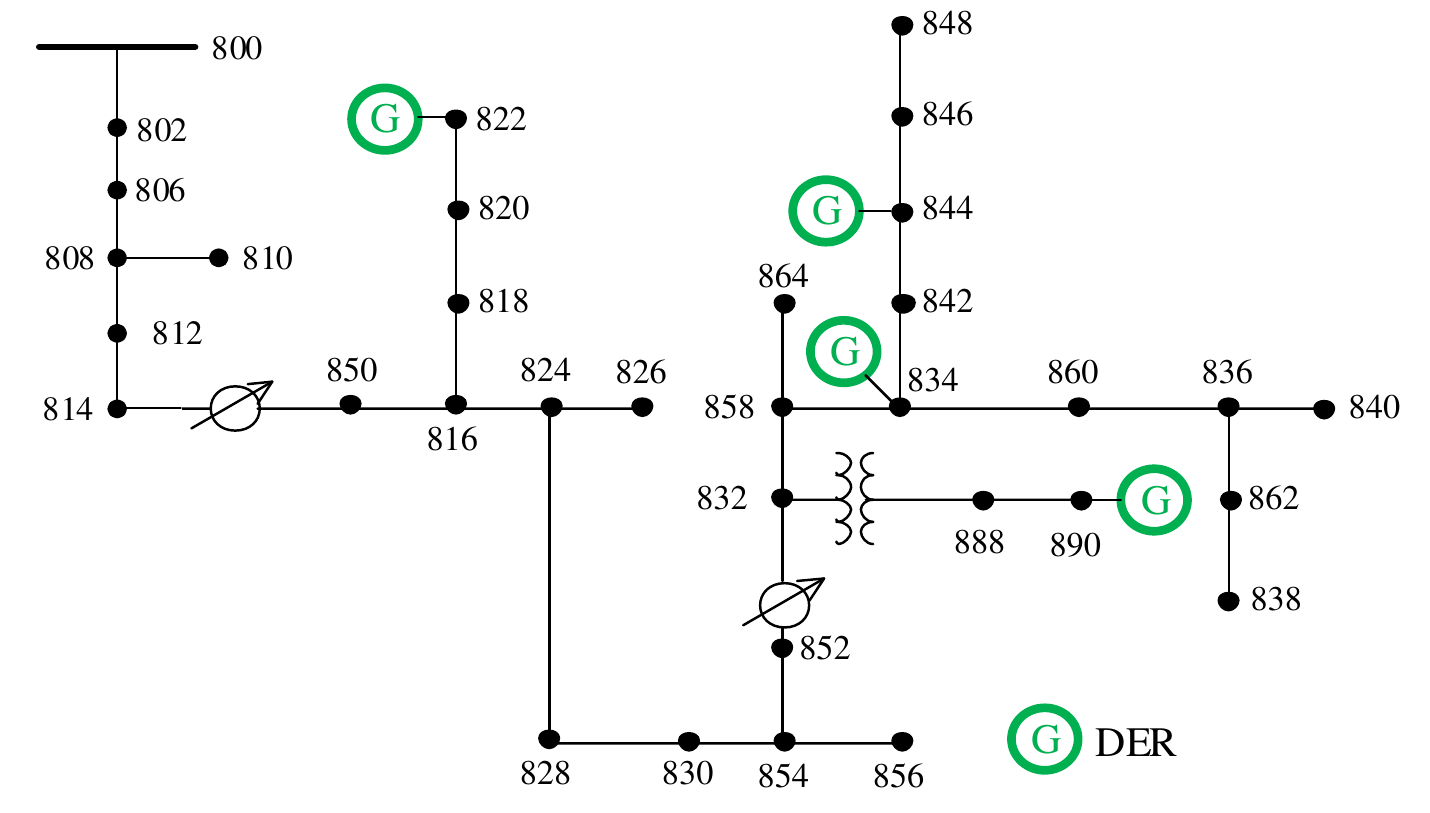}
    \vspace{-2.5ex}
    \caption{The modified IEEE 34-node distribution test system}
    \label{fig:34node}
    \vspace{-2ex}
\end{figure}

The proposed approach is implemented on the modified IEEE 34-node distribution test system. 
The IEEE 34-node system is an actual distribution system in Arizona, USA with a nominal voltage of 24.9 kV. It is characterized by being long and lightly loaded, and having two in-line regulators, an in-line transformer for a short 4.16 kV section, shunt capacitor banks, and unbalanced loading with constant current, power, and impedance models. For the detailed data of the IEEE 34-node system, the readers are referred to reference \cite{IEEEFEEDERS}. In this paper, this system has been modified by including three 3-phase DERs at nodes 844, 890, and 834, and a single-phase DER at phase 1 of node 822 as shown in Figure~\ref{fig:34node}. 

\begin{table}[]
    \centering
  \caption{Hyper-parameter Settings of the Proposed DRL Framework}
  \vspace{-2ex}
\begin{tabular}{|c|c|}
\hline
Hyper-parameters  & Values     \T \\ \hline \hline
Number of hidden layers & 2  \T\\
(both actor and critic networks) &  \\\hline
No. of neurons in hidden layers & 8, 8 \T \\ (both actor and critic networks)    &  \\ \hline
Learning rate of actor network & 0.001 \T   \\ \hline
Learning rate of critic network & 0.002 \T   \\ \hline
Reward discount factor   & 0.99   \T  \\ \hline
Activation function of output layer& Linear  \T \\ (critic network)  &\\ \hline
Activation function of output layer  & Sigmoid  \T \\(actor network)  &\\ \hline
Activation function of hidden layers & ReLU   \T   \\ (both actor and critic networks) &\\ \hline
Optimizer (both actor and critic networks)& Adam     \T     \\ \hline
Replay memory size  & 1500   \T        \\ \hline
Batch size  & 200    \T     \\ \hline
\end{tabular}
    \label{tab:hyper_parameters}
\end{table}
\subsection{Training of the proposed framework}
The training of the proposed framework for the modified IEEE 34-node system is performed for 1500 episodes. The parameters $\theta$ and $\phi$, respectively, of the main actor and critic networks are initialized with random values and the parameters $\theta_{targ}$ and $\phi_{targ}$ of the target networks are set equal to main network parameters. In each episode, the system state (i.e., a vector of maximum available reserves of each DER, the total reserve requested by the system operator, and reserve bid prices of each DER) is randomly initialized. 
\begin{figure}
    \centering
    \includegraphics[scale=0.55]{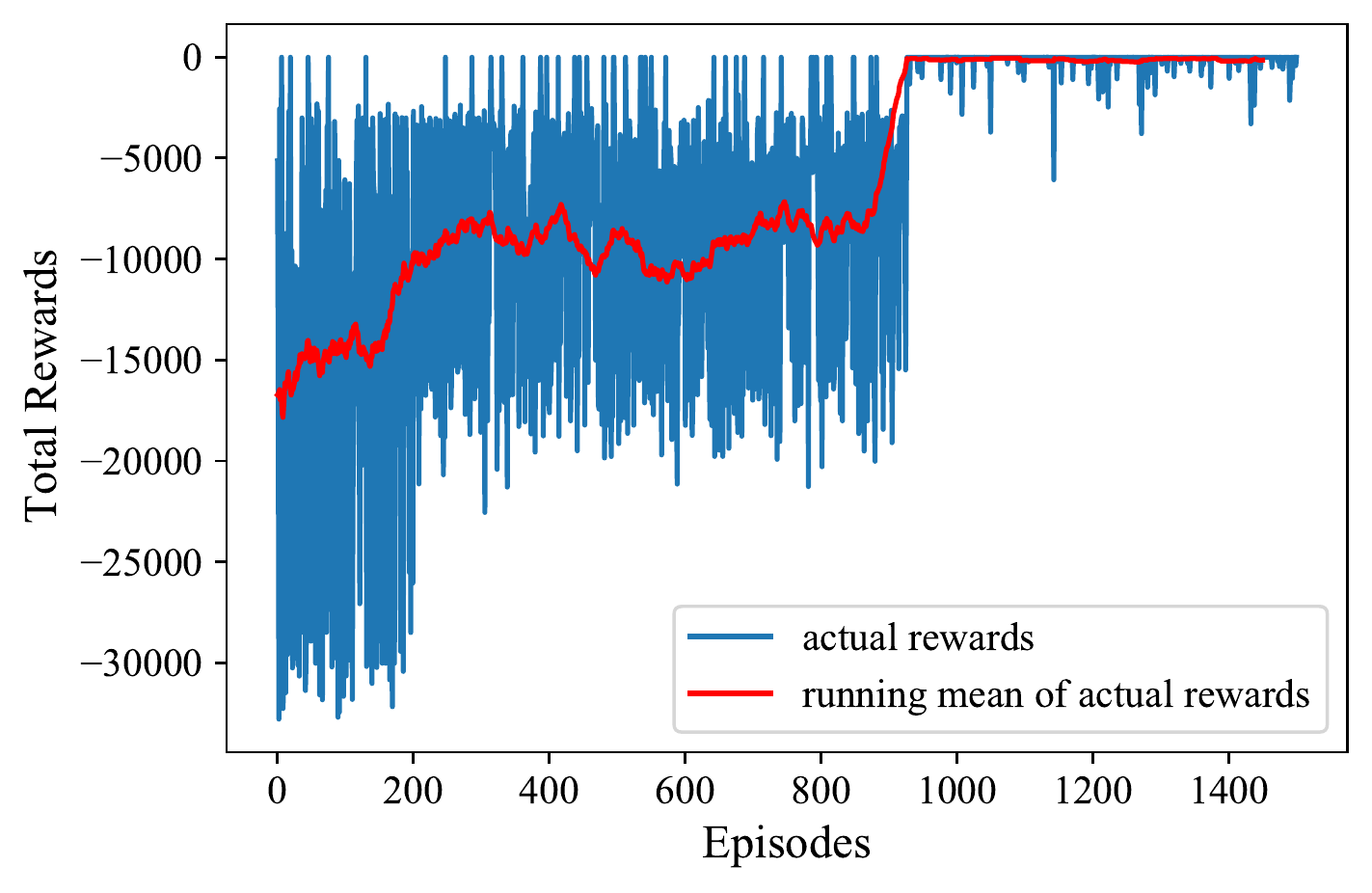}
    \vspace{-2ex}
    \caption{Total rewards of training episodes}
    \vspace{-3ex}
    \label{fig:reward_34}
\end{figure}
The total rewards are low up to nearly 1000 episodes but they increase after 1000 episodes and remain almost constant. Fig.~\ref{fig:reward_34} shows the actual rewards and running mean (50-episode window) of actual rewards as the episode progresses. It can be seen from the figure that as the number of episodes increases, the running mean of the total reward increases and saturates after nearly 1000 episodes.

\subsection{Testing, Implementation, and Comparison}
To test the effectiveness of the proposed framework for reserve optimization, two test cases are devised with different values of maximum available reserves, reserve bid prices, and total reserve requested by the system operator. Table~\ref{tab:test_param} shows the parameters of DERs for two test cases.  
\begin{table}[h!]
    \caption{DER Parameters for Two Test Cases}
    \vspace{-2ex}
    \centering
\begin{tabular}{|c|c|c|c|c|}
\hline
\multirow{2}{*}{DER-$i$} & \multicolumn{2}{c|}{Test   Case-I} & \multicolumn{2}{c|}{Test   Case-II} \T\\ \cline{2-5}  & $R_i^{max}$ (kW)      & $\pi_i$(\textcentoldstyle/kWh)      & $R_i^{max}$ (kW)      & $\pi_i$ (\textcentoldstyle/kWh)       \T \\ \hline \hline
1   & 200  & 10   & 100   & 12    \T   \\  \hline
2  & 200   & 12    & 80   & 10  \T \\ \hline
3  & 150 & 11    & 100  & 10   \T    \\ \hline
4  & 200   & 14  & 100& 12   \T   \\ \hline
\end{tabular}
\vspace{-2ex}
    \label{tab:test_param}
\end{table}
A vector of maximum available reserves of DERs, reserve bid prices of DERs, and the total reserve requested by the system operator is given as an input to the proposed framework and a vector of scheduled reserves is obtained as an output. 
For comparison purposes, results obtained using the proposed approach are compared with a capacity-based approach. In the capacity-based approach, reserves allocated for DERs are proportional to their available capacities of reserves. 

For Test Case-I, for 600 kW of reserve requested by the system operator for a particular scheduling period, scheduled reserves of DERs 1, 2, 3, and 4, respectively, using the proposed approach are 165.03 kW, 180.87 kW, 150 kW, and 104.10 kW; and that using the capacity-based approach are 160 kW, 160 kW, 120 kW, and 160 kW. Similarly, for Test Case-II, for 350 kW of reserve requested by the system operator for a particular scheduling period, scheduled reserves of DERs 1, 2, 3, and 4, respectively, using the proposed approach are 98.11 kW, 80 kW, 100 kW, and 71.89 kW; and that using the capacity-based approach are 92.10 kW, 73.70 kW, 92.10 kW, and 92.10 kW. These values are shown in Table~\ref{tab:reserves}.   

\begin{table}[]
    \caption{Schedules Reserves of DERs (in kW)}
    \vspace{-2ex}
    \centering
\begin{tabular}{|c|c|c|c|c|}
\hline
\multirow{2}{*}{DERs} & \multicolumn{2}{c|}{Test   Case-I} & \multicolumn{2}{c|}{Test   Case-II} \T\\ \cline{2-5}  & Proposed  & Capacity-based & Proposed  & Capacity-based \T\\
  & approach      &  approach    & approach      & approach \\ \hline \hline
1   & 165.03  & 160.00  & 98.11 & 92.10  \T   \\  \hline
2  & 180.87 & 160.00   & 80.00 & 73.70 \T \\ \hline
3  & 150.00 & 120.00    & 100.00  & 92.10  \T    \\ \hline
4  & 104.10  & 160.00  & 71.89  & 92.10  \T   \\ \hline
\end{tabular}
\vspace{-3ex}
    \label{tab:reserves}
\end{table}

After allocating reserves for Test Case-I as per Table~\ref{tab:reserves}, total cost of reserve, total power loss, and average voltage deviation, respectively, for the proposed approach are \$69.28, 114.18 kW, and 2.70\%; and that for the capacity-based approach are \$70.80, 121.15 kW, and 2.74\%. These values are shown in Table~\ref{tab:comparison_1}. Similarly, the results for Test Case-II are obtained, which are shown in Table~\ref{tab:comparison_2}. The comparison of results shows that the proposed approach performs better compared to the capacity-based approach.

\begin{table}
    \caption{Comparison of the Proposed Approach with Capacity-based Approach for Test Case-I}
    \vspace{-2ex}
    \centering
\begin{tabular}{|c|c|c|c|}
\hline
\multirow{2}{*}{Approaches} & \begin{tabular}[c]{@{}c@{}}Total reserve  \T \\ cost (\$/h)\end{tabular} & \begin{tabular}[c]{@{}c@{}}Total power\\ loss (kW)\end{tabular} & \begin{tabular}[c]{@{}c@{}}Average \\voltage\\ deviation\end{tabular} \T \\ \hline \hline
Proposed   approach   & 69.28 & 114.18 & 2.70 \%    \T  \\ \hline
Capacity-based   approach   & 70.80 & 121.15 & 2.74 \%   \T  \\ \hline
\end{tabular}
    \label{tab:comparison_1}
\end{table}

\begin{table}
    \caption{Comparison of the Proposed Approach with Capacity-based Approach for Test Case-II}
    \vspace{-2ex}
    \centering
\begin{tabular}{|c|c|c|c|}
\hline
\multirow{2}{*}{Approaches} & \begin{tabular}[c]{@{}c@{}}Total reserve  \T \\ cost (\$/h)\end{tabular} & \begin{tabular}[c]{@{}c@{}}Total power\\ loss (kW)\end{tabular} & \begin{tabular}[c]{@{}c@{}}Average \\voltage\\ deviation\end{tabular} \T \\ \hline \hline
Proposed   approach   & 38.40 & 172.60 & 2.64 \%    \T  \\ \hline
Capacity-based   approach    & 38.68 & 175.40 & 2.73 \%   \T  \\ \hline
\end{tabular}
\vspace{-3ex}
    \label{tab:comparison_2}
\end{table}


When the study is performed on a PC with 64-bit Intel i5 core, 3.15 GHz processor, 8 GB RAM, and Windows OS, the execution time of the proposed DRL framework is approximately 2 milli-seconds. 
\section{Conclusion}\label{conclusion}
This paper has proposed a DRL-based framework to utilize aggregated DERs for tertiary frequency regulation, and optimally schedule the requested reserves among the participating DERs. The proposed framework was formulated with a vector of available reserves and reserve prices submitted by DERs and the total reserve requested by the system operator as the state, and a vector of reserve schedules of each DER as the action. The reward function was designed in a way to minimize total reserve cost, network loss, and average voltage deviation. The deep deterministic policy gradient (DDPG), an algorithm based on the actor-critic method, was adopted to solve the optimization problem. In order to prevent catastrophic forgetting, a mechanism referred to as experience replay was used. Target networks were built for the stabilization of the update rule for training the main networks. Case studies were performed on the modified IEEE 34-node distribution system, and the results exhibit the effectiveness of the proposed framework to optimize the reserve with co-optimization of reserve cost by DERs, network power loss, and average voltage deviation in the network.     

\section*{Acknowledgement}
This paper is based upon work supported by the U.S. Department of Energy's Office of Energy Efficiency and Renewable Energy (EERE) under the Solar Energy Technologies Office Award Number DE-EE0009022. 

\bibliographystyle{IEEEtran}
\bibliography{References.bib}


\end{document}